\documentclass[a4paper,11pt]{article}
\usepackage{pos}
\usepackage{xcolor}

\usepackage{tikz}
\usetikzlibrary{arrows.meta}
\usetikzlibrary{decorations.markings}
\tikzstyle midarrow=[postaction={decorate,decoration={markings,
      mark=at position 0.53 with {\arrow{#1}},}}]

\title{Form factors for semileptonic $B\to\pi$, $B_s\to K$ and $B_s\to
  D_s$ decays}
\author*[a,b]{Jonathan Flynn}
\author*[a,c]{Ryan Hill}
\author[d,a,b]{Andreas J\"uttner}
\author[e]{Amarjit Soni}
\author[f]{J~Tobias Tsang}
\author[g]{Oliver Witzel}

\affiliation[a]{Physics \& Astronomy, University of Southampton,
  Southampton, SO17 1BJ, UK}
\affiliation[b]{STAG Research Centre, University of Southampton,
  Southampton SO17 1BJ, UK}
\affiliation[c]{DISCnet Centre for Doctoral Training, University
  of Southampton, Southampton SO17 1BJ, UK}
\affiliation[d]{CERN TH Division, 1211 Gen\`eve 23, Switzerland}
\affiliation[e]{Physics Department, Brookhaven National Laboratory,
  Upton, NY 11973, USA}
\affiliation[f]{CP3-Origins and IMADA, University of Southern
  Denmark, Campusvej 55, 5230 Odense M, Denmark}
\affiliation[g]{Center for Particle Physics Siegen, Theoretische Physik 1,
  Naturwissenschaftlich-Technische Fakult\"at,
  Universit\"at Siegen, 57068 Siegen, Germany}
\emailAdd{j.m.flynn@soton.ac.uk}
\emailAdd{r.c.hill@soton.ac.uk}

\abstract{We report on our determinations of $B\to \pi\ell\nu$,
  $B_s\to K \ell \nu$ and $B_s\to D_s \ell \nu$ semileptonic form
  factors. In addition we discuss the determination of $R$-ratios
  testing lepton-flavor universality and suggest an improved ratio.
  Our calculations are based on the set of 2+1 flavor domain-wall
  Iwasaki gauge field configurations generated by the RBC/UKQCD
  collaboration with three lattice spacings of $1/a = 1.78$, $2.38$,
  and $2.79\,\text{GeV}$. We use the relativistic heavy quark action
  for $b$ quarks and charm quarks are simulated with the M\"obius
  domain-wall fermion action.}

\FullConference{%
 The 38th International Symposium on Lattice Field Theory, LATTICE2021
  26th--30th July, 2021
  Zoom/Gather@Massachusetts Institute of Technology\\[1ex]
  {\rm Siegen preprint: SI--HEP--2021--36}
}

\def\w{\omega}
\def\qsqmax{{q^2_\text{max}}}

\def\gev{\,\text{Ge\hspace{-0.1em}V}}
\def\mev{\,\text{Me\hspace{-0.1em}V}}
\def\fm{\,\text{fm}}

\begin{document}
\maketitle

\section{Introduction}

Semileptonic decays of $B_{(s)}$ mesons play an important role in
testing and constraining the Standard Model (SM) of elementary
particle physics. Focusing on exclusive semileptonic decays, we report
on our work for $B\to\pi\ell\nu$, $B_s\to D_s\ell\nu$ and $B_s\to
K\ell\nu$ decays. Each of these processes can be described by two form
factors, $f_+$ and $f_0$, which parametrize the semileptonic decay
rate. For the semileptonic decay of pseudoscalar meson $B_{(s)}$ of
mass $M$ and momentum $p$ to pseudoscalar meson $P$ of mass $m$ and
momentum $k$, with $q=p-k$,
\begin{multline}
  \frac{d\Gamma(B_{(s)}{\to} P\ell\nu)}{dq^2} =\\
  \eta\frac{G_F^2 |V_{xb}|^2}{24\pi^3} \,
  \frac{(q^2{-}m_\ell^2)^2 |\vec k|}{(q^2)^2}
    \bigg[ \Big(1{+}\frac{m_\ell^2}{2q^2}\Big)
           \vec k^{\,2}\,|f_+(q^2)|^2
      +\frac{3m_\ell^2}{8q^2}
      \frac{(M^2{-}m)^2}{M^2}\,|f_0(q^2)|^2
      \bigg],
    \label{eq:difftl-decay-rate}
\end{multline}
where $m_\ell$ is the mass of the outgoing charged lepton $\ell$ and
$\eta$ is an isospin factor. The form factors $f_+$ and $f_0$ appear
in the decomposition
\begin{equation}
  \langle P(k) |\mathcal{V}^\mu(0) | B_{(s)}(p) \rangle
    %= f_+(q^2)\bigg(p^\mu +k^\mu -\frac{M^2_{B_s} - M^2_P}{q^2}q^\mu\bigg)
    = 2\,f_+(q^2)\bigg(p^\mu -\frac{p\cdot q}{q^2}q^\mu\bigg)
    + f_0(q^2)\frac{M^2-m^2}{q^2}q^\mu,
\end{equation}
where $\mathcal V^\mu = \bar x\gamma^\mu b$, with $x=u$ or $c$.

Compared to our earlier results for $B\to\pi$ and $B_s\to K$
decays~\cite{Flynn:2015mha}, we have added calculations of $B_s\to
D_s$ form factors and have an additional, third, lattice spacing. With
results for $B_s\to K$ and $B_s\to D_s$ from the same ensembles, we
will be able to compute the ratio of partially integrated decay rates
(minus CKM factors) in the region $q^2\geq 7\gev^2$ for the two decays
and combine with recent LHCb results~\cite{Aaij:2020nvo} to determine
$|V_{ub}/V_{cb}|$. We also consider $R$ ratios of branching fractions
with $\tau$ leptons in the final state to those with light final-state
leptons, sensitive to violations of lepton-flavor universality (for
$B\to D^{(*)}\ell\nu$ there is tension between Standard
  Model predictions and experimental results for
  $R(D^{(*)})$~\cite{Lees:2012xj,Lees:2013uzd,% babar
  Huschle:2015rga,Hirose:2016wfn,Hirose:2017dxl,
  Abdesselam:2019dgh,Belle:2019rba,% belle
  Aaij:2015yra,Aaij:2017uff,Aaij:2017deq% lhcb
}). We
propose a modified ratio with smaller uncertainty when evaluated using
lattice-determined form factors.

\section{Lattice calculation}

We use a subset of six RBC/UKQCD $2{+}1$-flavor domain-wall fermion
(DWF) and Iwasaki gauge field ensembles with three lattice spacings $a
\sim 0.11$, $0.08$, $0.07\fm$, and pion masses spanning $267\mev <
M_\pi < 433\mev$. The ensembles are listed in
Table~\ref{tab:ensembles}. Light and strange quarks are simulated with
the Shamir DWF action with $M_5=1.8$. Lattice spacings are determined
from combined RBC/UKQCD
analyses~\cite{Blum:2014tka,Boyle:2017jwu,Boyle:2018knm}. Our
calculations are described briefly below; for more details
see~\cite{Flynn:2015mha}.
\begin{table}
\[
  \begin{array}{cccccccccccc}
  & L & T & L_s &  a^{-1}\!/\!\gev & am_l^\text{sea} & am_s^\text{sea} 
  & M_\pi/\!\mev & \text{\# cfgs} & \text{\# sources}\\\hline
\text{C1}&24&64&16 &1.785 & 0.005 & 0.040 & 340 &1636 & 1\\
\text{C2}&24&64&16 &1.785 & 0.010 & 0.040 & 433 &1419 & 1\\[1.2ex]
\text{M1}&32&64&16 &2.383 & 0.004 & 0.030 & 302 &628  & 2\\
\text{M2}&32&64&16 &2.383 & 0.006 & 0.030 & 362 &889  & 2\\
\text{M3}&32&64&16 &2.383 & 0.008 & 0.030 & 411 &544  & 2\\[1.2ex]
\text{F1S}&48&96&12 &2.785  & 0.002144 & 0.02144 & 267 &98 & 24
  \end{array}
  \]
  \caption{Ensembles used for the simulations reported
    here~\cite{Allton:2008pn,Aoki:2010dy,Blum:2014tka,Boyle:2017jwu}.
    $am_l^\text{sea}$ and $am_s^\text{sea}$ are the sea light and
    strange quark masses and $M_\pi$ is the unitary pion mass. Valence
    strange quarks are near their physical mass, with the mistuning
    accounted for in our systematic errors.}
  \label{tab:ensembles}
\end{table}
  
Bottom quarks are simulated with the relativistic heavy quark (RHQ),
action, which is the Columbia variant~\cite{Christ:2006us,Lin:2006ur}
of the Fermilab heavy-quark action~\cite{ElKhadra:1996mp}, with three
nonperturbatively-tuned parameters
$(m_0a,c_P,\zeta)$~\cite{Aoki:2012xaa}. A new tuning was performed for
this analysis. Charm quarks are simulated with the M\"obius DWF action
with
$M_5=1.6$~\cite{Boyle:2016imm,Boyle:2017jwu,Boyle:2017kli,Boyle:2018knm}.
We use three masses below $m_c^\text{phys}$ on the C ensembles and two
masses which bracket $m_c^\text{phys}$ on M and F. Light and strange
quarks have point sources; while the $b$ and $c$ quarks use
Gauss-smeared sources and point or smeared sinks.

Renormalized, $\mathcal{V}_\mu$, and lattice, $V_\mu$, currents are
related by the `partially nonperturbative'
procedure~\cite{Hashimoto:1999yp,ElKhadra:2001rv}, using
\begin{equation}
    \langle P | \mathcal{V}_\mu | B_s \rangle = Z^{bx}_{V_{\mu}}\langle P |
  V_\mu | B_s\rangle,
\end{equation}
with
$Z^{bx}_{V_\mu} = \rho^{bx}_{V_\mu}\sqrt{Z^{xx}_VZ^{bb}_V}$ and
\begin{equation}
 V_0 = V_0^{0} + c_t^{3} V_0^{3} + c_t^{4} V_0^{4}, \qquad
 V_i = V_i^{0}+ c_s^{1} V_i^{1} + c_s^{2} V_i^{2}
           + c_s^{3} V_i^{3} + c_s^{4} V_i^{4}.
\end{equation}
Here, $\rho^{bx}_{V_\mu}$ and the coefficients $c_{t,s}^n$ of the
$O(a)$ current-improvement operators are computed perturbatively at
one-loop~\cite{CLehnerPT}, while $Z^{bb}_V$ is computed
nonperturbatively from the forward matrix element
\begin{equation}
  Z_V^{bb}\langle B_{(s)}|V_0(0)|B_{(s)} \rangle = 2M
\end{equation}
and $Z_V^{xx}$ is computed nonperturbatively using the relation
$Z_V^{xx} = Z_A^{xx} + O(am_{\text{res}})$ for DWF
fermions~\cite{Aoki:2010dy} (for our systematic error analysis for
$B_s\to D_s$ decays, we also compare to using
$Z_V^{cc}=Z_A^{ll}$~\cite{Boyle:2017jwu}).

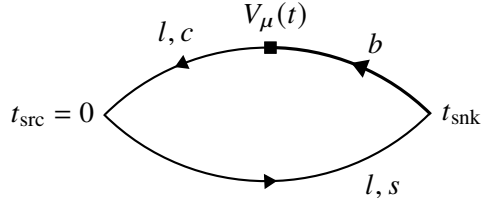
\begin{figure}
\begin{center}
  \begin{tikzpicture}[x=0.1\textwidth,y=0.1\textwidth]
\useasboundingbox (-0.46,-0.7) rectangle (3.3,0.85);
%\draw[red] (-0.46,-0.7) rectangle (3.3,0.85);
\draw[black,thick,midarrow={Triangle}] (1.425,0.585) arc(90:135:2)
node [pos=1,left] {$t_\text{src}=0$} node[pos=0.35,above left]{$l,c$};
\draw[black,very thick,midarrow={Triangle}] (2.845,0) arc(45:90:2)
node [pos=0,right] {$t_\text{snk}$} node[pos=0.5,above right]{$b$};
\draw[black,thick,midarrow={Triangle}] (0,0) arc(-135:-45:2)
node[pos=0.75,below right]{$l,s$};
\draw[fill=black] (1.4,0.535) rectangle (1.5,0.635)
node[above]{$V_\mu(t)$};
\end{tikzpicture}
\end{center}
\caption{Three-point correlator used in form-factor determinations.}
\label{fig:3ptcorr}
\end{figure}

\begin{figure}
  \begin{center}
    \includegraphics[width=0.67\textwidth,
      trim=15 10 20 35,clip=true]{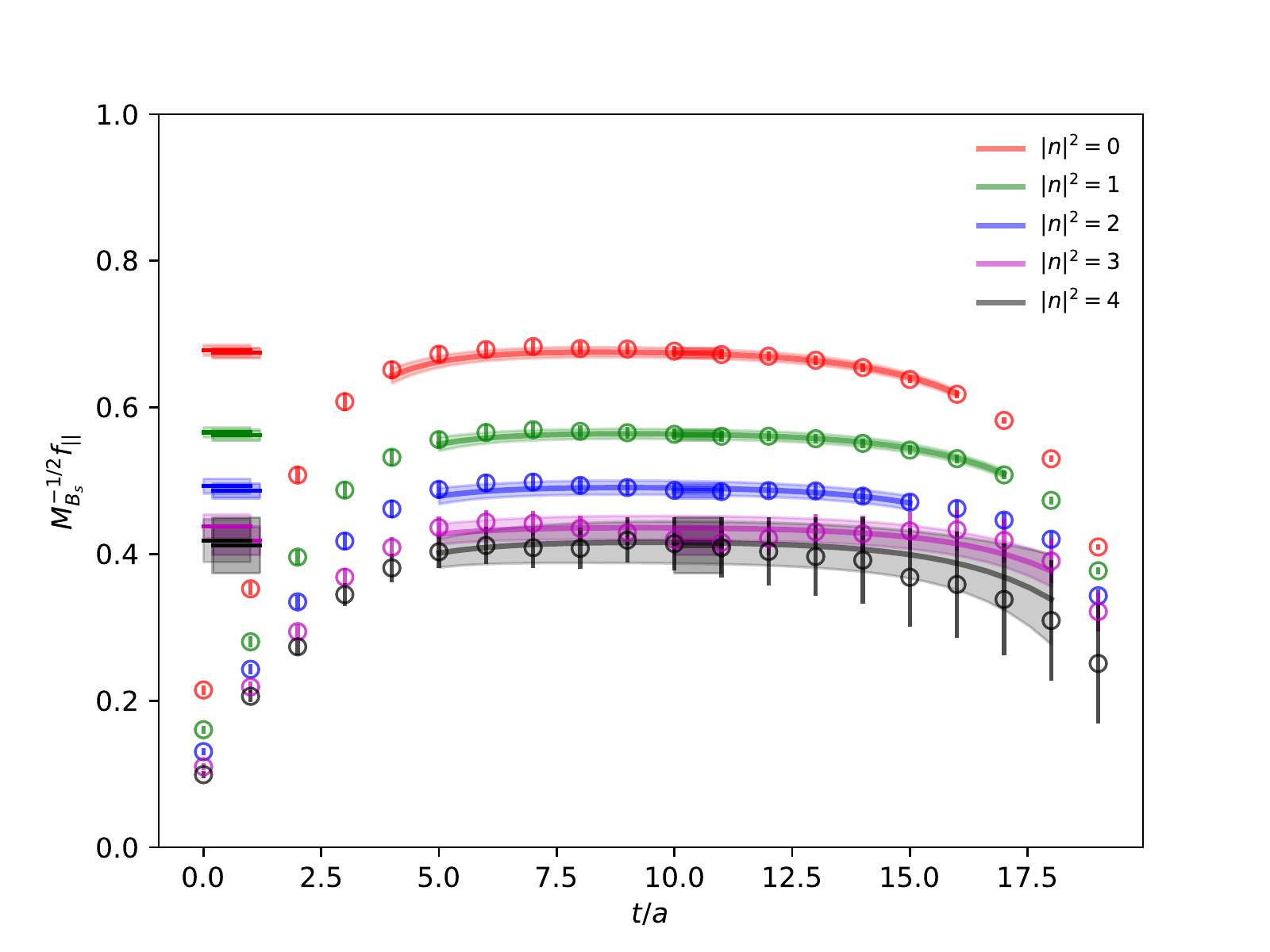}
  \end{center}
  \caption{Extraction of $f_\|$ for $B_s\to K$ on the coarse ensemble
    C1 from the ratio $R_{3,0}$. The different colors denote
    different three-momenta $2\pi\vec n/L$ injected at the current,
    labelled by $n^2$. The plot shows a ground-state-only fit together
    with a fit over an extended range of times for each momentum once
    excited state terms are included for the current matrix elements
    in the numerator of $R_{3,0}$. The horizontal bars near the left
    axis show the values for $f_\|$ from the ground-only and
    from the excited-state fits. }
  \label{fig:R-ratio}
\end{figure}

To extract the form factors we first calculate the matrix elements
\begin{equation}
  f_\parallel(E) =\frac{\langle P | \mathcal{V}^0(0) |
  B_{(s)}\rangle}{\sqrt{2M}},\qquad 
  f_\perp(E) = \frac{\langle P | \mathcal{V}^i(0) |
    B_{(s)}\rangle}{k^i\sqrt{2M}},
\end{equation}
with a $B_{(s)}$ meson at rest, where $E$ is the energy of the
outgoing pseudoscalar meson, from which we determine
\begin{align}
f_0(q^2) &= \frac{\sqrt{2M}}{M^2-m^2}\left[ (M -
  E)f_\parallel(E) + (E^2 - m^2)f_\perp(E)\right],\\
f_+(q^2) &= \frac{1}{\sqrt{2M}}\left[f_\parallel(E) + (M -
  E)f_\perp(E)\right].
\end{align}
To find $f_\|$ and $f_\perp$, we evaluate a correlator ratio
\begin{equation}
  R_{3,\mu}(t,t_\text{snk}, \vec{k}) =
  \frac{C_{3, \mu}(t,t_\text{snk},\vec{k})}
  {\sqrt{C_2^{P}(t,\vec{k}) C_2^{B_{(s)}}(t_\text{snk}{-}t,\vec{0})}}
  \sqrt{\frac{2E}{e^{-E t -M(t_\text{snk}-t)}}},
\end{equation}
where $C_2^{P,B_{(s)}}$ are two-point correlators and $C_{3,\mu}$ is
the three-point correlator shown schematically in
figure~\ref{fig:3ptcorr}.
For large time separations between source, sink and current insertion,
we obtain
\begin{equation}
  f_\parallel^\text{bare}(\vec{k}) =
  \lim_{0\ll t\ll t_\text{snk}} R_{3,0}(t,t_\text{snk},\vec{k}), \qquad
  f_\perp^\text{bare}(\vec{k}) =
  \lim_{0\ll t\ll t_\text{snk}} \frac{1}{p^i_P} R_{3,i}(t,t_\text{snk},\vec{k}).
\end{equation}
Figure~\ref{fig:R-ratio} illustrates the determination of $f_\|$ for
$B_s\to K$ on the coarse, C1, ensemble.

\begin{figure}
  \begin{center}
    \includegraphics[width=.494\textwidth]{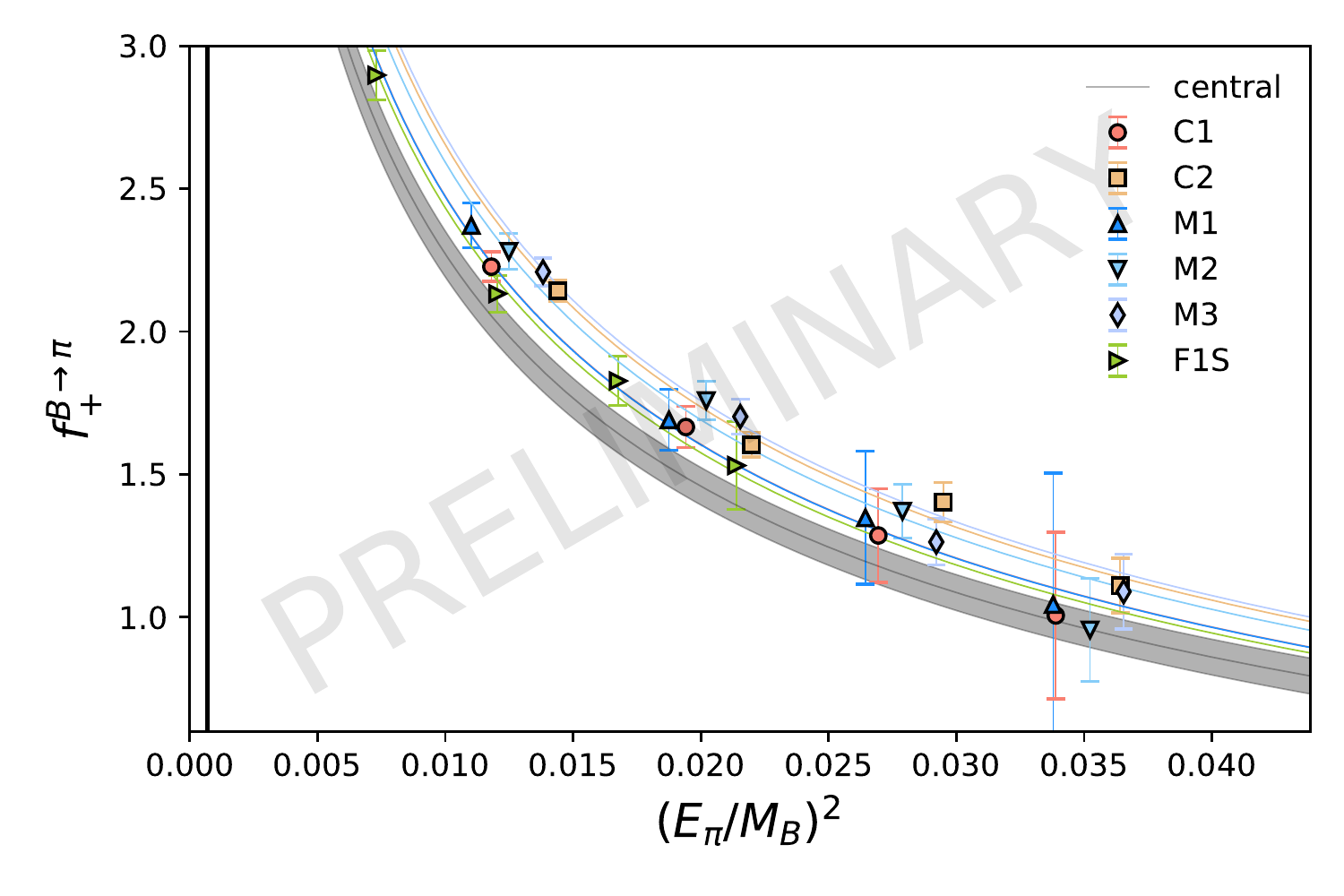}
    \includegraphics[width=.494\textwidth]{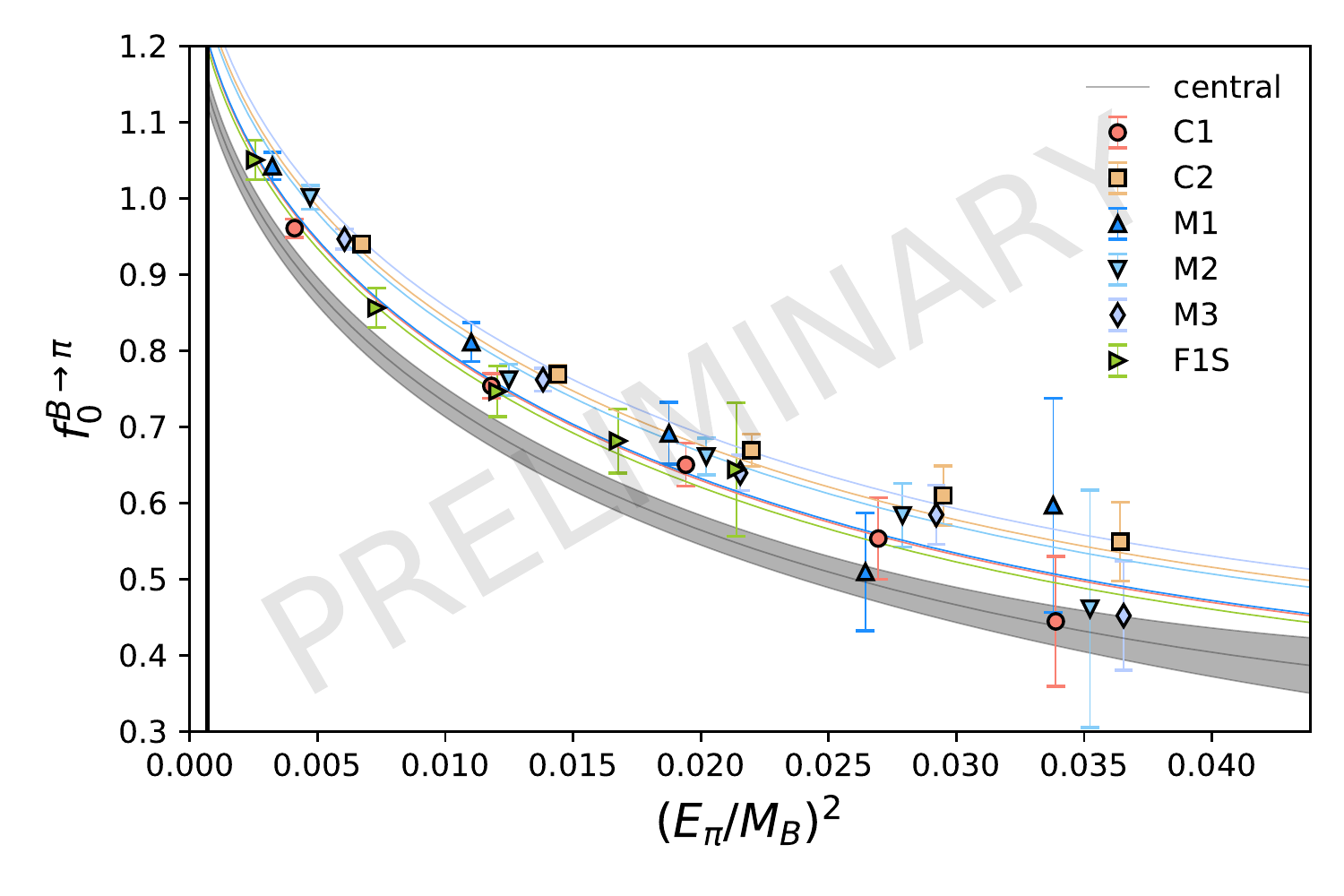}
  \end{center}
\caption{Chiral-continuum extrapolation for the $B\to\pi$ form
  factors $f_+$ (left) and $f_0$ (right). The colored data points
  show the underlying data. The colored lines show the result of the
  fit evaluated at the parameters of the respective ensembles. The
  grey bands display the form factors in the chiral-continuum
  limit and the associated statistical uncertainty.}
\label{fig:Btopi-chiral-ctm-fit}
\end{figure}

\begin{figure}
  \begin{center}
    \includegraphics[width=.494\textwidth]{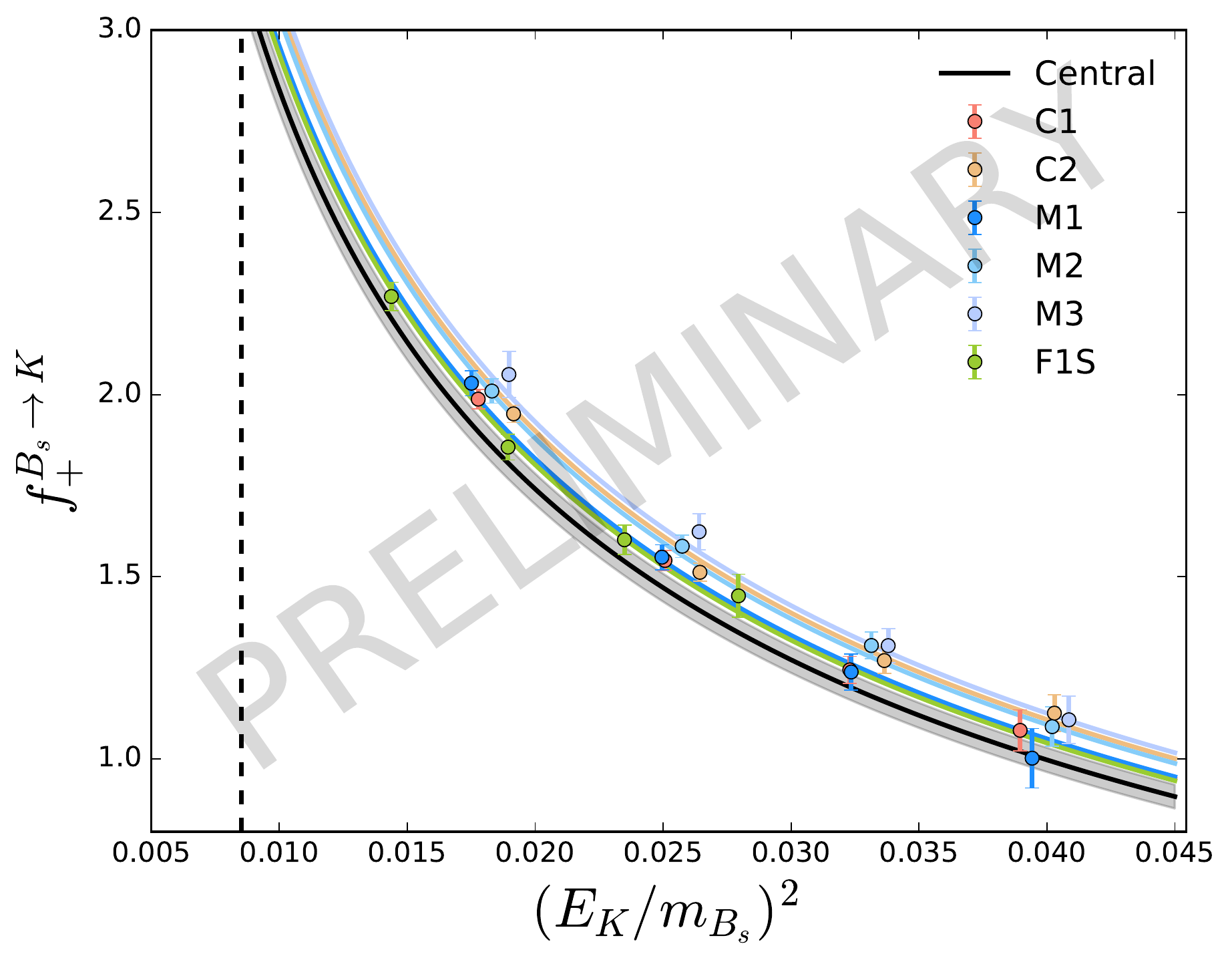}
    \includegraphics[width=.494\textwidth]{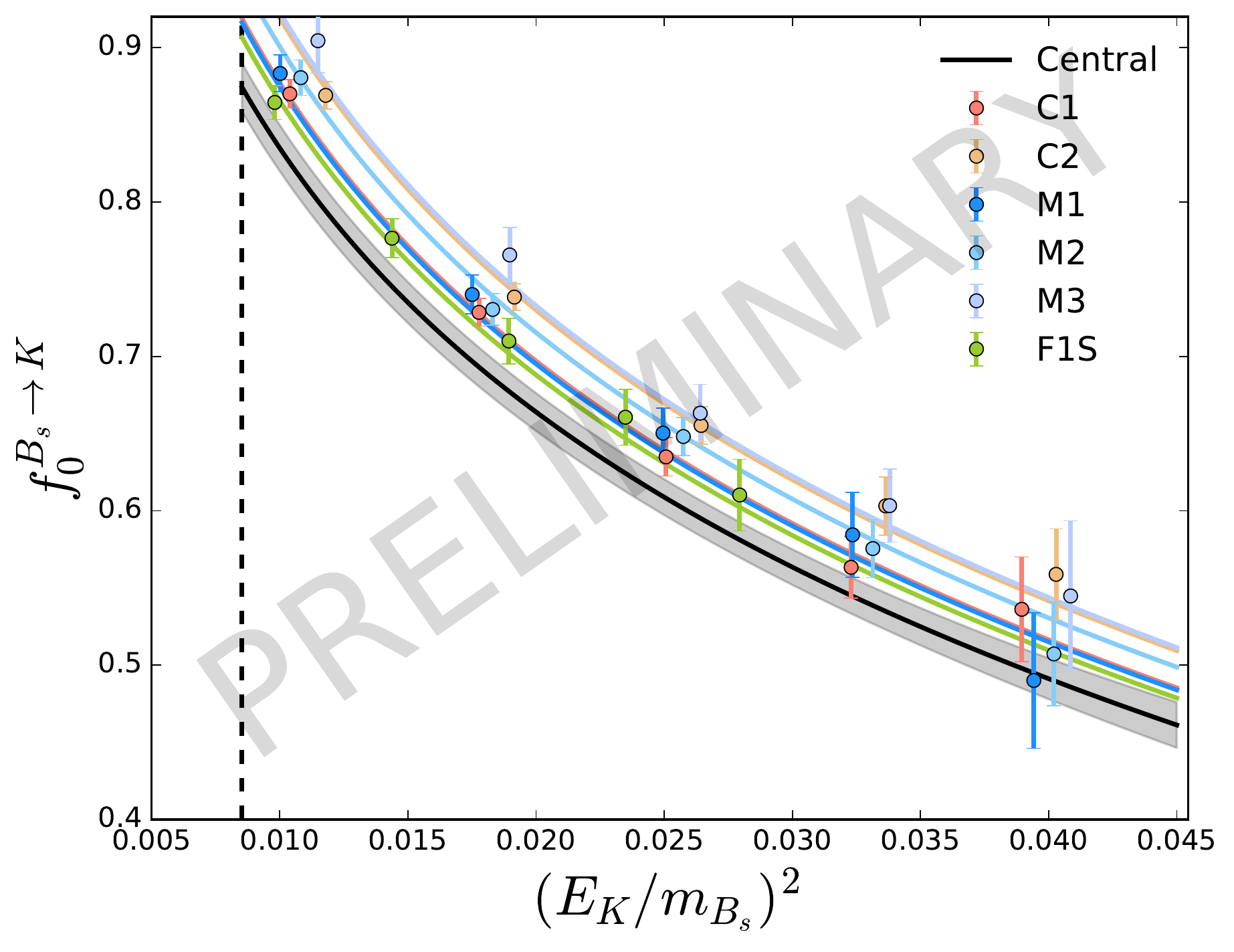}
  \end{center}
\caption{Chiral-continuum extrapolation for the $B_s\to K$ form
  factors $f_+$ (left) and $f_0$ (right). The colored data points
  show the underlying data. The colored lines show the result of the
  fit evaluated at the parameters of the respective ensembles. The
  grey bands display the form factors in the chiral-continuum
  limit and the associated statistical uncertainty.}
\label{fig:BstoK-chiral-ctm-fit}
\end{figure}

For $B_s\to K$ and $B\to\pi$ we extrapolate the renormalized lattice
form factors to vanishing lattice spacing and to the physical
light-quark mass, and interpolate in the kaon(pion) energy, using
next-to-leading order SU(2) heavy-meson chiral perturbation theory
(HM$\chi$PT) in the ``hard-kaon(pion)''
limit~\cite{Flynn:2008tg,Bijnens:2010ws,Becirevic:2002sc}. The
function we use, with $P$ denoting kaon or pion, is
%\begin{equation}
  \begin{multline}
  \label{eq:f-HMChPT}
    f^{B_{(s)}\to P}(M_{\pi}, E_P, a^2) =
  \frac{\Lambda}{E_P+\Delta}
  \left[c_0\bigg(1+
    \frac{\delta f(M^\text{sea}_\pi)-\delta
      f(M^\text{phys}_\pi)}{(4\pi f_\pi)^2}
    \bigg)\right.\\
  \left.{}+ c_1 \frac{\Delta M_\pi^2}{\Lambda^2}  
  + c_2\frac{E_P}{\Lambda}
  + c_3\frac{E_P^2}{\Lambda^2} 
  + c_4(a\Lambda)^2 \right],
  \end{multline}
%\end{equation}
where $M_\pi^\text{sea}$ is the simulated pion mass on a given
ensemble, $M_\pi^\text{phys}$ is the physical pion mass, $\Delta
M_\pi^2 = (M_{\pi}^\text{sea})^2 - (M_{\pi}^\text{phys})^2$ and
$\Lambda=1\gev$ is the renormalization scale appearing in the one-loop
chiral logarithm in $\delta f$, and is also used as a dimensionful
scale to render the fit coefficients dimensionless. $\Delta =
M_{B^*}-M_{B_{(s)}}$ and the $B^*$ is a $\bar bu$ flavor state with
$J^P=1^-$ for $f_+$, or $J^P=0^+$ for $f_0$. For $f_+$ this is the
vector meson $B^*$ with mass $M_{B^*} =
5.32470(22)\gev$~\cite{Tanabashi:2018oca}, while for $f_0$ there is a
theoretical estimate for the $0^+$ state,
$M_{B^*(0^+)}=5.63\gev$~\cite{Bardeen:2003kt}. The term $\delta f$
also contains an estimate for finite volume effects.
Figure~\ref{fig:Btopi-chiral-ctm-fit} shows the fit for $B\to\pi$ and
Figure~\ref{fig:BstoK-chiral-ctm-fit} shows the fit for $B_s\to K$.

\begin{figure}
  \begin{center}
    \includegraphics[width=.494\textwidth]{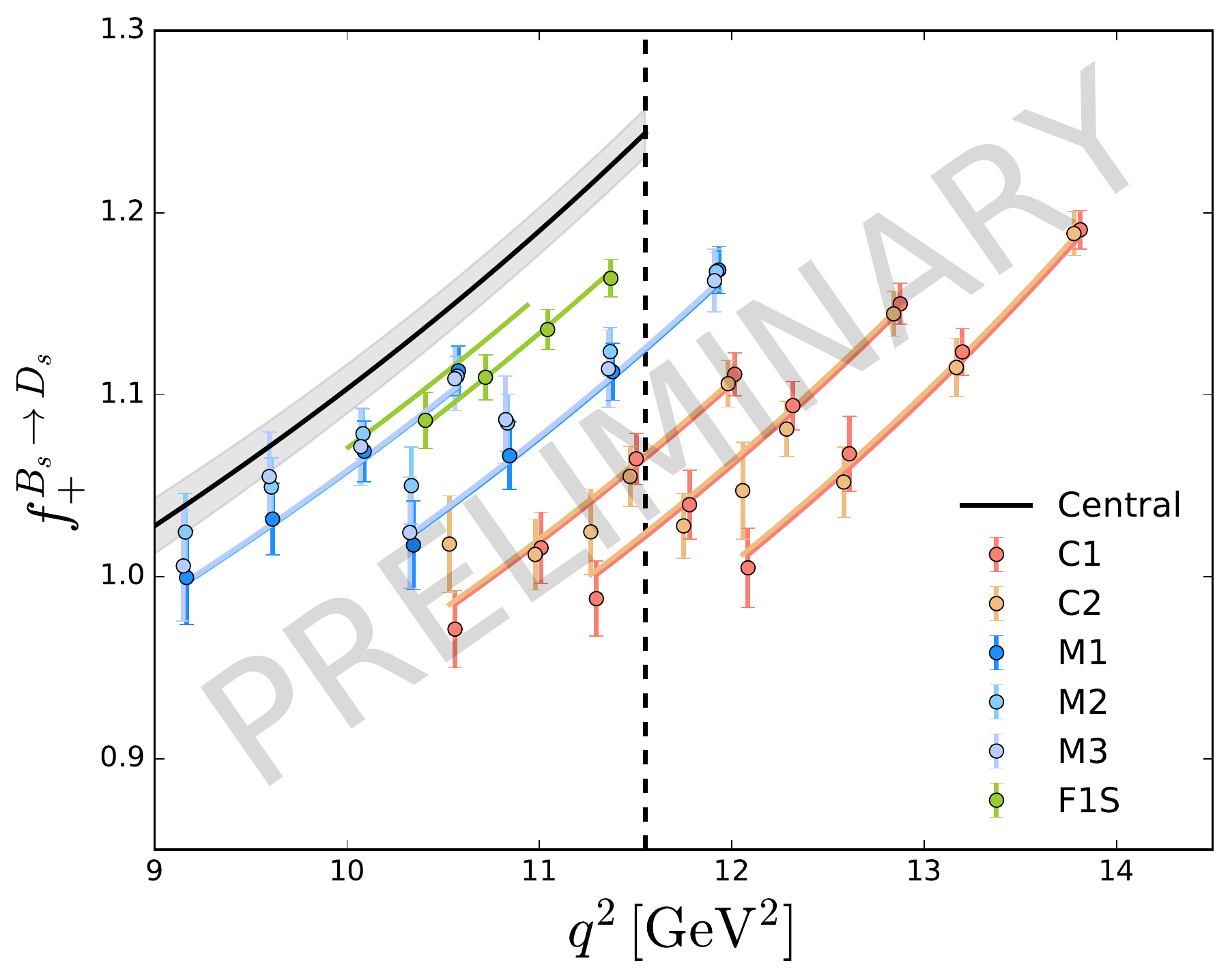}
    \includegraphics[width=.494\textwidth]{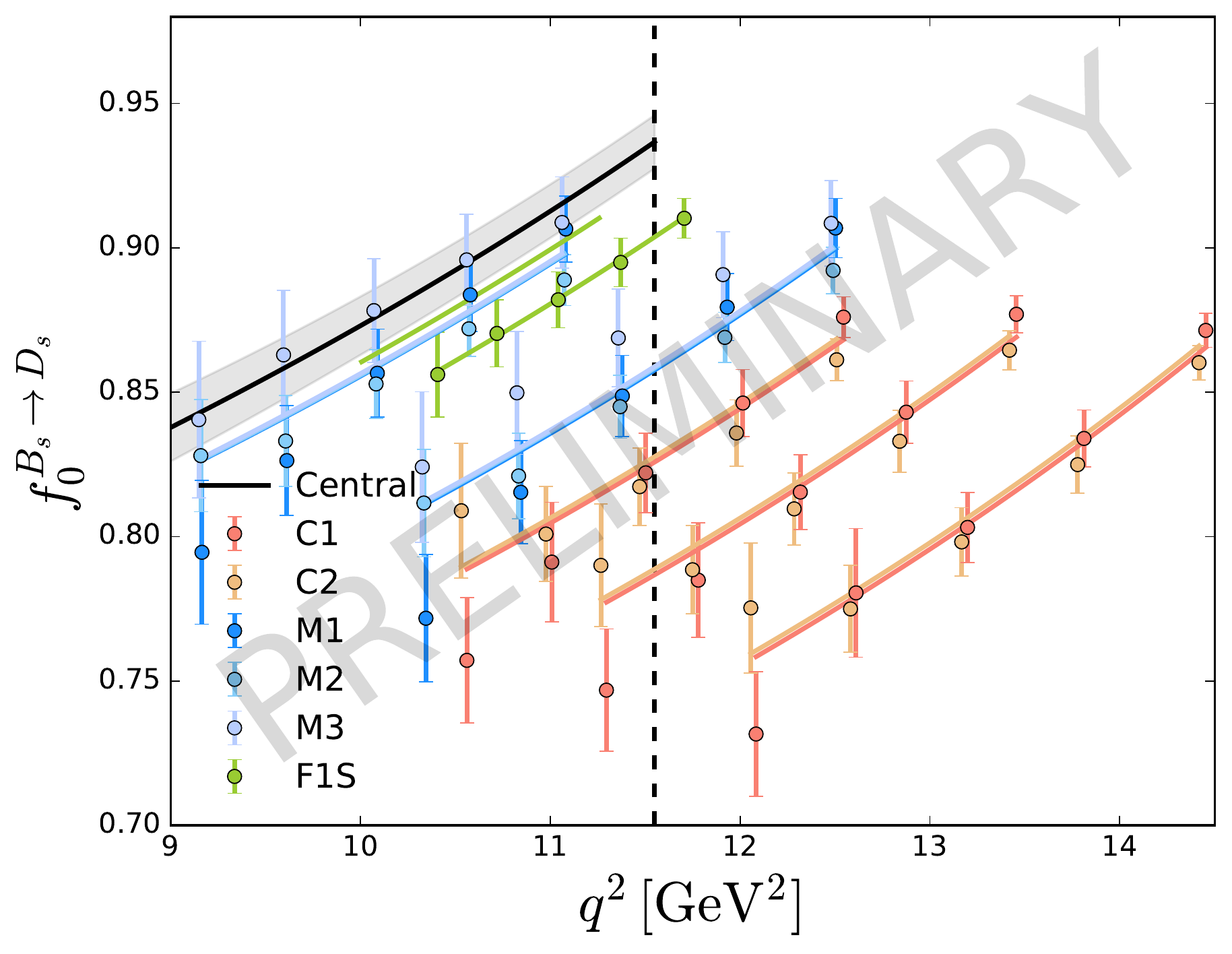}
  \end{center}
\caption{Chiral-continuum extrapolation for the $B_s\to D_s$ form factors.}
\label{fig:BstoDs-chiral-ctm-fit}
\end{figure}

For $B_s\to D_s$ form factors, we combine a chiral-continuum fit with
an extra-/inter-polation in the charm mass with a fit form
\begin{equation}
  f(q^2\!,a,M_\pi,M_{D_s}) =
  \bigg[c_0 +\! \sum_{j=1}^{n_{D_s}} \!c_{1j}\,
    h\Big(\frac{M_{D_s}}{\Lambda}\Big)^j\! + c_2 (a\Lambda)^2
    \bigg]
  P_{a,b}(q^2/M_{B_s}^2),
\end{equation}
where
\begin{equation}
  %% h\big(\frac{M_{D_s}}{\Lambda}\big) =
  %% \frac{\Lambda}{M_{D_s}}-\frac{\Lambda}{M_{D_s}^\text{phys}}
  h\Big(\frac{M_{D_s}}{\Lambda}\Big) =
  \frac{M_{D_s}}{\Lambda}-\frac{M_{D_s}^\text{phys}}{\Lambda}
  \qquad\text{and}\qquad
P_{a,b}(x) = \frac{1+\sum_{i=1}^a a_i x^i}{1+\sum_{i=1}^b b_i x^i}.
\end{equation}
Figure~\ref{fig:BstoDs-chiral-ctm-fit} shows the fit. We use only one
of the charm masses for F1S in this fit because of the strong
correlations between results for the two charm masses on that
ensemble.

\section{$z$-fits}

\begin{figure}
  \begin{center}
    \includegraphics[width=.494\textwidth]{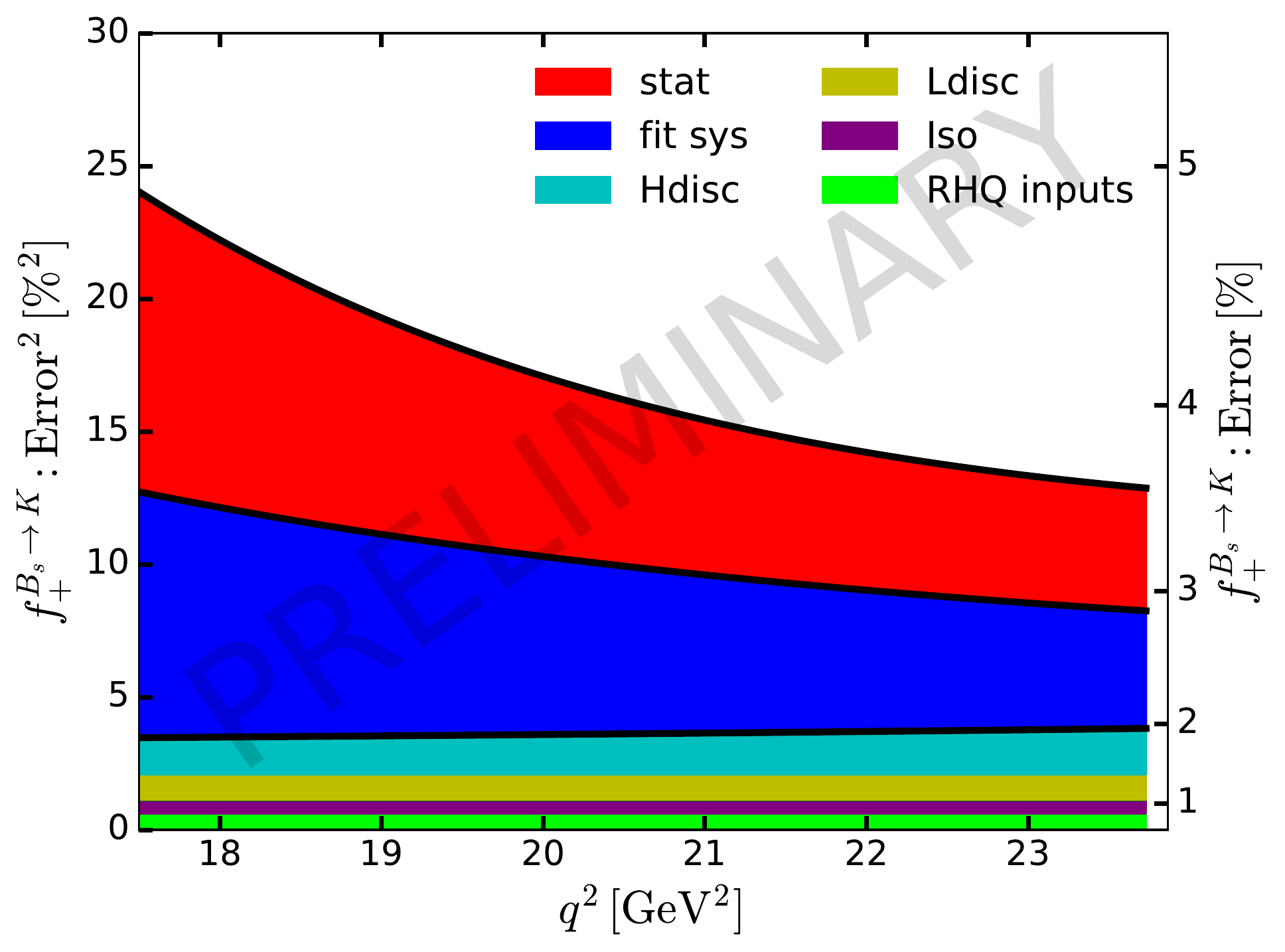}
    \includegraphics[width=.494\textwidth]{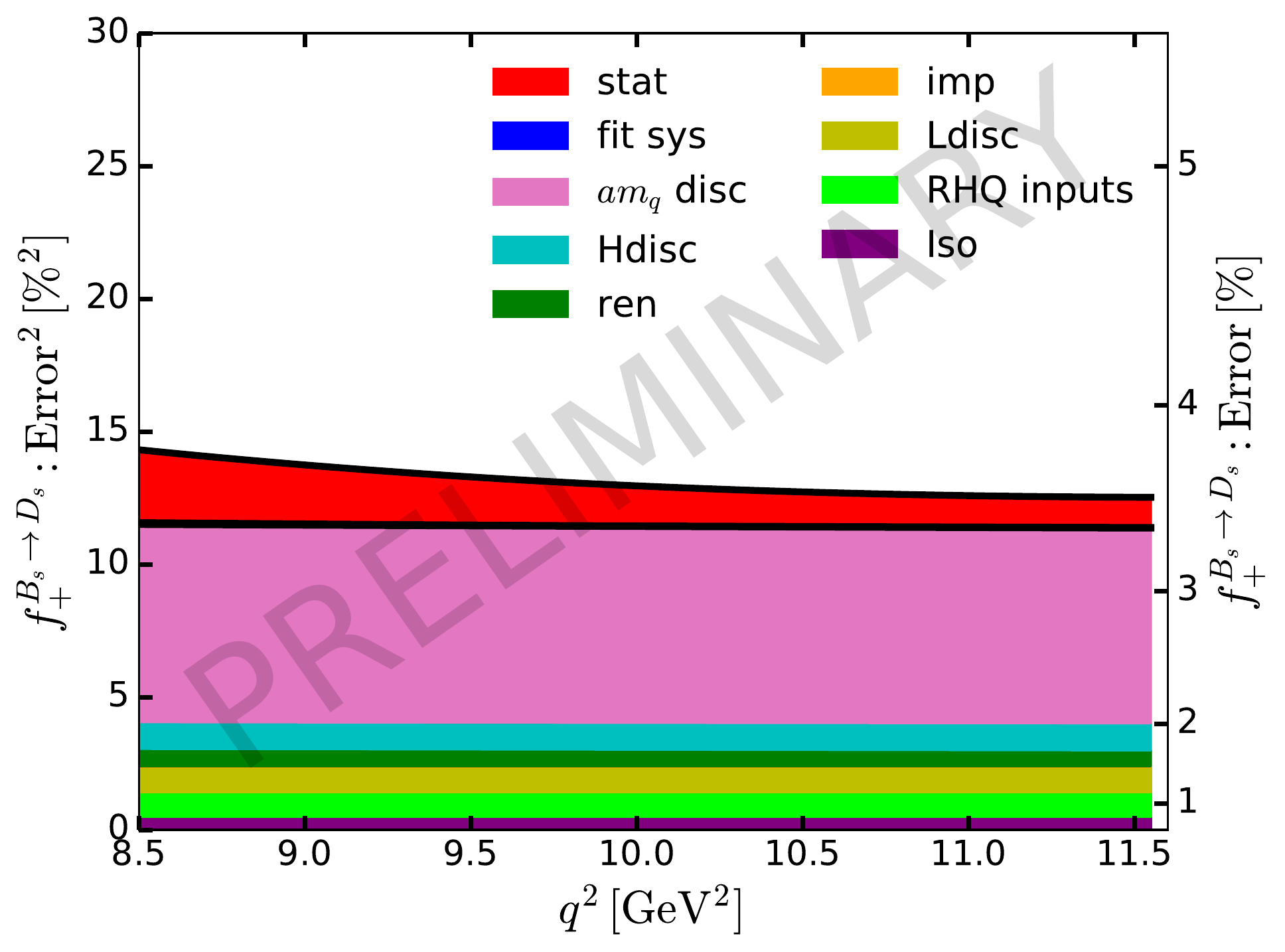}
  \end{center}
\caption{Cumulative error budgets for statistical and systematic
  errors for $f_+$ for $B_s\to K$ and $B_s\to D_s$. The $q^2$ ranges
  of the plots correspond to the ranges in which we generate synthetic
  data points for subsequent $z$-fits. The plots are for squared
  percentage errors, but an additional scale on the right shows the
  corresponding percentage error. The legends label those errors
  visible on the plots, but other error sources with sub-percent
  effects were considered. For $B_s\to K$, statistical errors (red)
  and systematic errors from the chiral-continuum extrapolation (blue)
  dominate. For $B_s\to D_s$, the dominant error (pink) is from
  discretization for the charm quark.}
\label{fig:error-budgets}
\end{figure}

\begin{figure}
  \begin{center}
    \includegraphics[width=.494\textwidth]{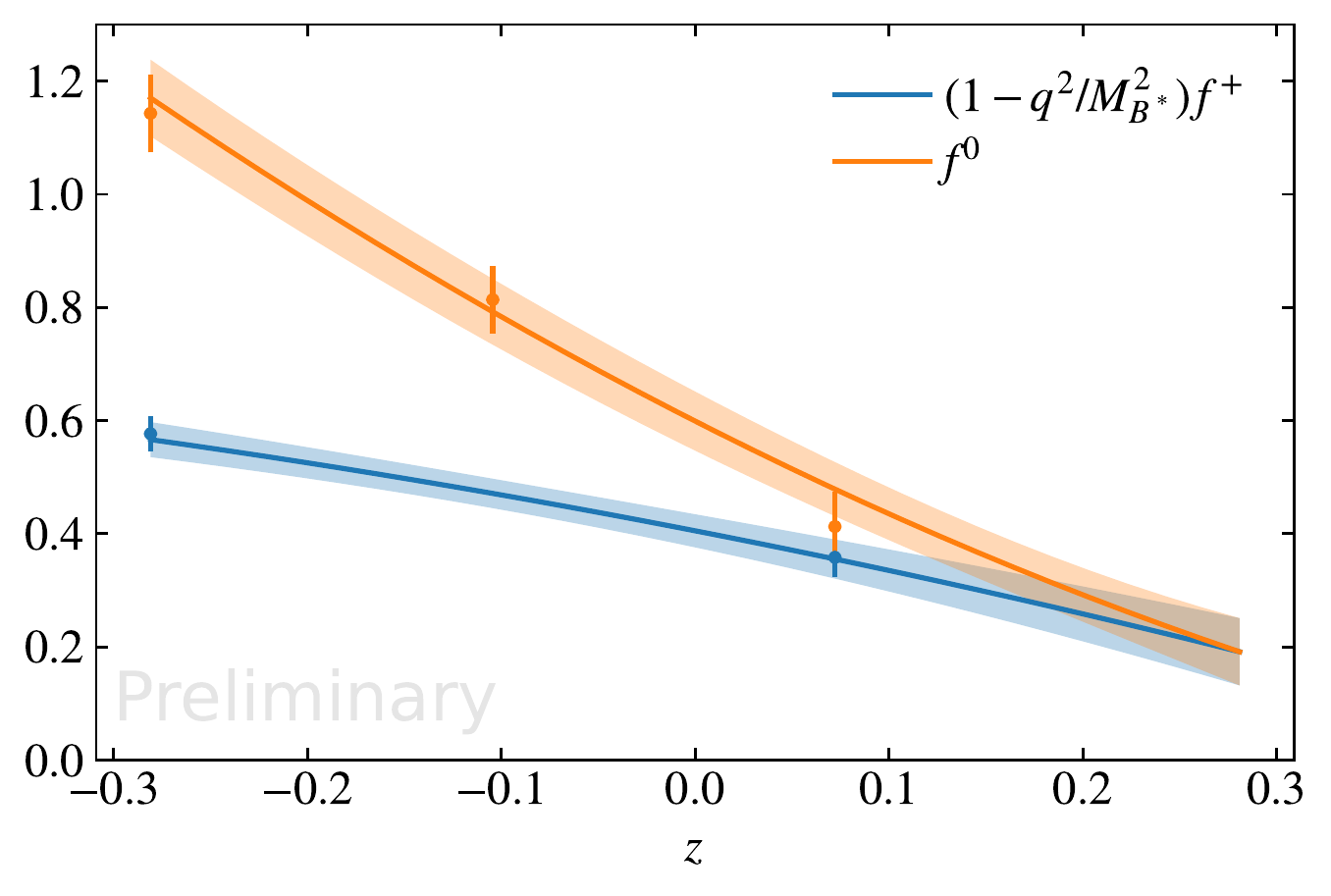}
    \includegraphics[width=.494\textwidth]{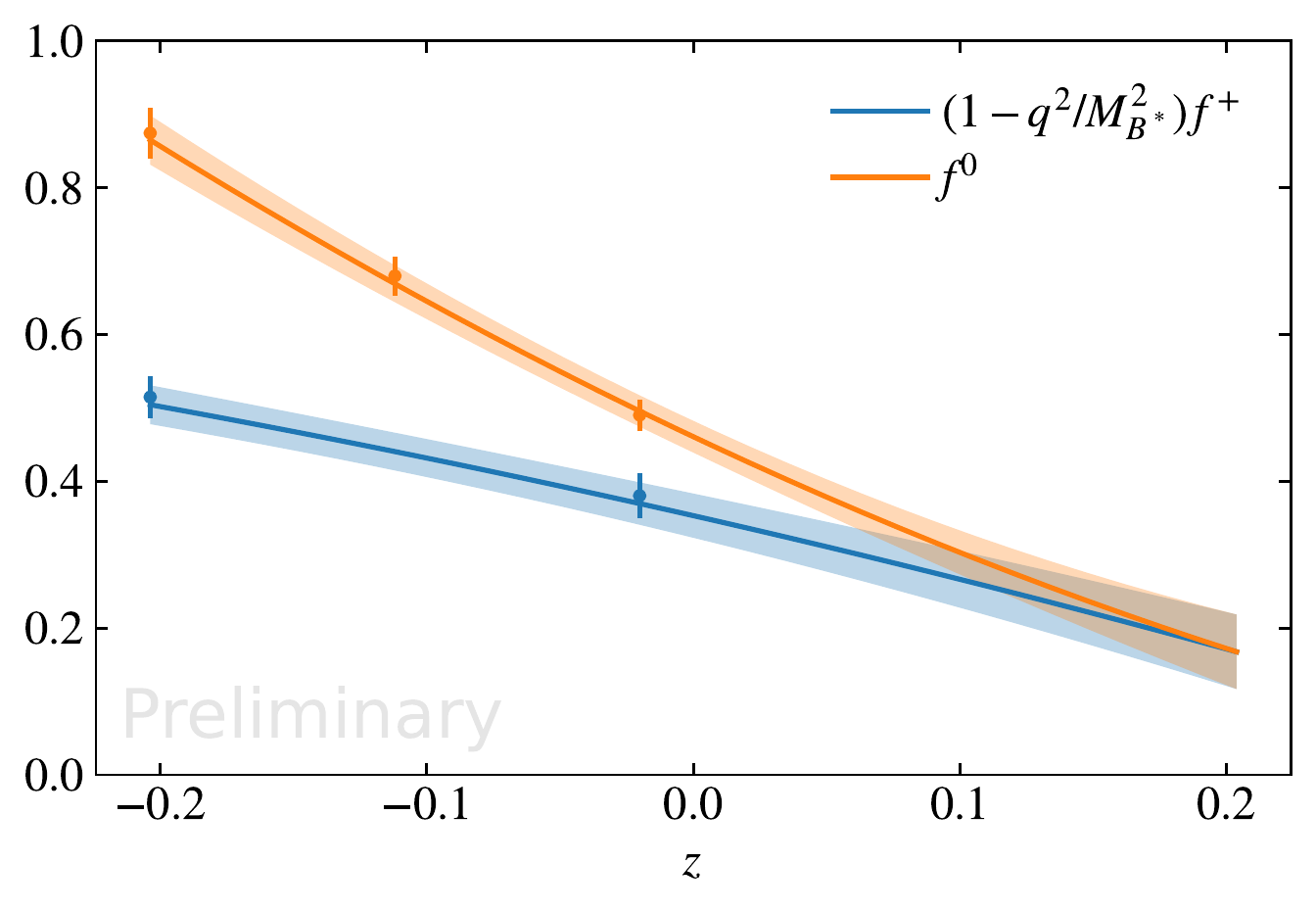}
  \end{center}
\caption{BCL $z$-fits for $B\to\pi$ (left) and $B_s\to K$ (right) form
factors.}
\label{fig:z-fits}
\end{figure}

After extrapolating our results to the continuum and physical masses,
our strategy is to generate synthetic data points for the form
factors, with all errors included, which can then be used in standard
$z$-fits to extrapolate over the full $q^2$ range for the physical
form factors. In figure~\ref{fig:error-budgets}, we illustrate the
cumulative statistical plus systematic error budgets for the $f_+$
form factor for $B_s\to K$ and $B_s\to D_s$ decays.
Figure~\ref{fig:z-fits} shows results for $z$-fits for $B\to\pi$ and
$B_s\to K$ form factors. These are Bourrely-Caprini-Lellouch (BCL)
fits~\cite{Bourrely:2008za}, where we have included the $1^-$ $B^*$
vector meson pole for fitting $f_+$ and no pole for $f_0$ in both
cases.

\section{Ratios for testing lepton flavor universality}

Ratios of decay rates or partially integrated decay rates with tau
leptons in the final state to those with light leptons in the final
state are of great interest in looking for violations of the
universality of lepton couplings in the Standard Model. For the
semileptonic decays considered here, such a ratio is $R(P)$ given by
\begin{equation}
  \label{eq:R}
      R(P)= \frac{\int_{m_\tau^2}^{q^2_\text{max}}dq^2\,
    \frac{d\Gamma(B_{(s)}\to P\tau\bar\nu_\tau)}{dq^2}}
    {\int_{m_\ell^2}^{q^2_\text{max}}dq^2\,
      \frac{d\Gamma(B_{(s)} \to P\ell\bar\nu_\ell)}{dq^2}},
\end{equation}
where $\ell$ in the denominator can be $\mu$ or $e$. We have
considered modifying this ratio to look for a sharper test of lepton
flavor universality (see discussion in~\cite{Atwood:1991ka} on
optimising observables). In particular, we try to reduce the
uncertainty in the ratio coming from uncertainties in the form factors
taken from our lattice simulations. To this end, we consider the
modified ratio
\begin{equation}
  \label{eq:Rnew}
    R^\text{new}(P) = \frac{\int_{q^2_\text{min}}^\qsqmax dq^2
     \frac{d\Gamma(B_{(s)}\to P\tau\bar\nu_\tau)}{dq^2}}
     {\int_{q^2_\text{min}}^\qsqmax dq^2
     \frac{\w_\tau(q^2)}{\w_\ell(q^2)}
     \frac{d\Gamma(B_{(s)}\to P\ell\bar\nu_\ell)}{dq^2}} ,
\end{equation}
following the recipe already applied to $B_{(s)}\to V$ decays (with a
vector meson instead of a pseudoscalar meson in the final state) by
Isidori and Sumensari~\cite{Isidori:2020eyd}. The ingredients are
\begin{itemize}
  \item Use a common integration range for numerator and denominator,
    with $q^2_\text{min} \geq
    m_\tau^2$~\cite{Freytsis:2015qca,Bernlochner:2016bci,Flynn:2020nmk}.
  \item Reweight the integrand in the denominator to make the
    contributions from the vector form factor the same in numerator
    and denominator.
\end{itemize}
Write the differential decay rate from
equation~(\ref{eq:difftl-decay-rate}) in the form
\begin{equation}
    \frac{d\Gamma(B_{(s)}{\to}P\ell\nu)}{dq^2} =
    \Phi(q^2) \w_\ell(q^2) \big[ F_V^2 + (F_S^\ell)^2\big],
\end{equation}
where now $\ell$ can be any lepton flavor, with
\begin{align}
  \Phi(q^2) &= \eta\frac{G_F^2 |V_{xb}|^2}{24\pi^3}
            |\vec k|,\\
  \w_\ell(q^2) &= \bigg(1-\frac{m_\ell^2}{q^2}\bigg)^2
            \bigg(1+\frac{m_\ell^2}{2q^2}\bigg),\\
  F_V^2 &= \vec k^2 |f_+(q^2)|^2,\\
  (F_S^\ell)^2 &= \frac34 \frac{m_\ell^2}{m_\ell^2+2q^2}
              \frac{(M^2-m^2)^2}{M^2}\, |f_0(q^2)|^2.
\end{align}
If we drop the scalar contribution, $(F_S^\ell)^2$, in the
denominator, with $\ell = \mu$ or $e$ again, then relying on
$m_\ell^2/2q^2 \leq m_\mu^2/2m_\tau^2 = 0.002$ for the light leptons,
we expect in the Standard Model,
\begin{equation}
    R^\text{new,SM}(P) = 1 + \frac{\int_{q^2_\text{min}}^\qsqmax dq^2\,
   \Phi(q^2) \w_\tau(q^2) (F_S^\tau)^2}
  {\int_{q^2_\text{min}}^\qsqmax dq^2\,
    \Phi(q^2) \w_\tau(q^2) F_V^2}.
\end{equation}
Evaluating the new ratio from equation~(\ref{eq:Rnew}) using the $z$-fit
for our lattice form factors, we find a reduced uncertainty compared
to evaluating the original ratio in equation~(\ref{eq:R}). It would be
interesting to see how the ratios compare if evaluated using
experimental differential decay rates.

\acknowledgments We thank our RBC/UKQCD collaborators for helpful discussions
and suggestions. Computations used resources provided by the USQCD
Collaboration, funded by the Office of Science of the US~Department of Energy
and by the \href{http://www.archer.ac.uk}{ARCHER} UK National Supercomputing
Service, as well as computers at Columbia University and Brookhaven National
Laboratory. We used gauge field configurations generated on the DiRAC Blue
Gene~Q system at the University of Edinburgh, part of the DiRAC Facility, funded
by BIS National E-infrastructure grant ST/K000411/1 and STFC grants
ST/H008845/1, ST/K005804/1 and ST/K005790/1.  The project leading to this
application has received funding from the European Union's Horizon 2020 research
and innovation programme under the Marie Sk{\l}odowska-Curie grant agreement No
894103.  This project has received funding from Marie Sk{\l}odowska-Curie grant
659322 (EU Horizon 2020), the European Research Council grant 279757 (EU
FP7/2007--2013), STFC grants ST/P000711/1 and ST/T000775/1. RH was supported by
the DISCnet Centre for Doctoral Training (STFC grant ST/P006760/1). AS was
supported in part by US DOE contract DE--SC0012704. JTT acknowledges support
from the Independent Research Fund Denmark, Research Project~1, grant
8021--00122. No new experimental data was generated.
  
\bibliographystyle{JHEP-jmf-arxiv}
\bibliography{B_meson}

\end{document}